\documentclass[aps,prl,twocolumn,floatfix,superscriptaddress]{revtex4}

\usepackage{graphicx}
\usepackage{dcolumn}
\usepackage{bm}

\begin{document}


\title{Atom-in-jellium equations of state for nitrogen, oxygen, and fluorine}


\author{Thomas~Lockard}
\author{Marius~Millot}
\affiliation{%
   Lawrence Livermore National Laboratory,
   7000 East Avenue, Livermore, California 94551, USA
}
\author{Burkhard~Militzer}
\affiliation{%
   University of California, Berkeley, California 94720, USA
}
\author{Sebastien~Hamel}
\author{Lorin~X.~Benedict}
\author{Philip~A.~Sterne}
\author{Damian~C.~Swift}
\affiliation{%
   Lawrence Livermore National Laboratory,
   7000 East Avenue, Livermore, California 94551, USA
}

\begin{abstract}
Equations of state (EOS) calculated from a computationally efficient 
atom-in-jellium treatment of the electronic structure
have recently been shown to be consistent with more rigorous
path integral Monte Carlo (PIMC) and quantum molecular dynamics (QMD) simulations
of metals in the warm dense matter regime.
Here we apply the atom-in-jellium model to predict wide-ranging EOS
for the cryogenic liquid elements nitrogen, oxygen, and fluorine.
The principal Hugoniots for these substances were surprisingly consistent with
available shock data and Thomas-Fermi (TF) EOS for very high pressures, 
and exhibited systematic variations from TF associated with shell ionization
effects, in good agreement with PIMC, though deviating from QMD and experiment
in the molecular regime.
The new EOS are accurate much higher in pressure than previous 
widely-used models for nitrogen and oxygen in particular,
and should allow much better predictions for oxides and nitrides
in the liquid, vapor, and plasma regime, where these have previously been
constructed as mixtures containing the older elemental EOS.
\end{abstract}


\maketitle

\section{Introduction}
Many of the most common planet-forming substances contain oxygen,
including silica and silicates, oxides such as alumina, MgO, and FeO, 
carbonates, water, and CO$_2$.
Some contain nitrogen, in particular ammonia.
Equations of state (EOS) used to understand planetary impacts and 
giant planets and exoplanets are typically constructed in the warm, dense matter
regime by mixing elemental EOS.
EOS for constituents mixed in this way should be reasonably accurate
at atomic volumes and temperatures corresponding to high pressure
compression of the corresponding cryogenic liquids.
The third second-row diatomic cryogenic liquid, fluorine,
is of practical interest as a component of SF$_6$,
used in dielectric breakdown switches in which it forms a plasma during operation,
and of LiF,
which is widely used in high pressure experimental
studies as an optical window or tamper.
Fluorine is also a constituent of various polymers and chemical explosives,
and is a component of `flibe,' a possible coolant 
and in-situ breeder of tritium in thermonuclear reactors,
where it would be subjected to heating and compression.

Shock wave data, usually used to calibrate high pressure EOS,
are limited in range for the cryogenic liquids oxygen and nitrogen,
and non-existent for fluorine.
Widely-used semi-empirical EOS from the {\sc sesame} and {\sc leos} libraries 
\cite{sesame,leos} are thus limited in data for calibration or validation.
The {\sc sesame} EOS for oxygen and nitrogen were constructed only for the
molecular regime, using a model of molecular vibrations (MV) for the 
ion-thermal energy \cite{Kerley1980}.
While tabulated to $10^8$\,K for use in wide-ranged applications,
they are not valid for temperatures over $10^4$\,K,
as they do not include the effects of molecular dissociation and ionization.
The {\sc sesame} EOS for fluorine \cite{Crockett2006} was constructed using 
a different procedure than other EOS in the {\sc sesame} library,
largely because of the lack of shock data.
An accurate but narrow-range experimentally-derived, thermophysical EOS was used to constrain
a Mie-Gr\"uneisen model for the ions, blended into Thomas-Fermi-Dirac (TFD)
calculations at high compressions.
TFD was used also for the electron-thermal energy, as in most {\sc sesame} EOS.

Recently, the rigorous but computationally expensive techniques of
path integral Monte Carlo (PIMC) \cite{pimc} 
and quantum molecular dynamics (QMD) \cite{qmd}
have shown that all-electron average-atom calculations,
in which the ions surrounding the atom are simplified as a
uniform `jellium'
\cite{Liberman1979}, can give accurate predictions of electronic
excitation contributions to the EOS of dense plasmas 
\cite{Benedict2014,Driver2017}.
We have found further that, for a set of well-studied reference elements, 
when atom-in-jellium electronic states are
combined with a perturbative approach to estimate
the ion-thermal contribution to the EOS \cite{Liberman1990},
they give reasonably accurate predictions of the complete EOS
in the liquid phase as well as the dense plasma regime
\cite{Swift2018,Swift2019}.


In the work reported here, 
we apply these same atom-in-jellium techniques to predict the EOS
of nitrogen, oxygen, and fluorine, and compare the results against previous
work, including combined QMD and PIMC studies \cite{Driver2016},
and shock wave experiments 
\cite{Zubarev1962,Dick1970,Nellis1980,Nellis1991,Trunin2008,Mochalov2010,Hamilton1988},
which have been reported
for both nitrogen and oxygen but not for fluorine.
We pay particular attention to the prediction of the shock Hugoniot for
initial states in the cryogenic liquid.
While the atom-in-jellium calculations are inaccurate where molecular
bonding occurs,
we find that a correction to the atom-in-jellium energy
in the initial state allows us to make accurate predictions of the
shock Hugoniot for final states in the warm dense matter regime
of interest for astrophysical impacts and other applications of high energy density.

\section{Atom-in-jellium calculations}
The cryogenic liquids are an interesting test of
electronic structure calculations of EOS.
The atom-in-jellium model was originally expected to be suitable only for
close-packed metals, as it represents the electron distribution
with spherical symmetry and does not capture the relative orientation
of neighboring atoms, smearing their charge into the uniform background jellium
outside the Wigner-Seitz sphere \cite{Liberman1979}.
However, atom-in-jellium calculations were subsequently found to be 
similarly accurate for the EOS of non-close-packed metals,
and even for carbon and silicon \cite{Swift2018}, 
despite the importance of directional
bonding in these elements near ambient conditions.

The diatomic cryogenic liquids are notable in that saturated interatomic
bonds and van der Waals forces are essential to their behavior at low
pressures and temperatures; these behaviors are not captured by the 
present atom-in-jellium model.
The atom-in-jellium EOS are thus not expected to be accurate at low pressures,
and it is therefore of interest to know how closely they match
existing data at higher pressures asnd temperatures, 
where directional chemical bonding becomes less important.
Additionally, we are interested in EOS predictions that can be made
at even higher pressures, where the EOS is currently unconstrained by
experimental data,
as well as comparisons with EOS predicted using other approaches
such as the Mie-Gr\"uneisen or Thomas-Fermi (TF) models \cite{tf}.

The average-atom calculations reported here
were performed using the same prescription as in our previous
study \cite{Swift2018}.
For each element, atom-in-jellium calculations were made over a range
and density of tabulation suitable for a general-purpose EOS:
mass density $\rho$ from
$10^{-4}$ to $10^3\rho_0$ with 20 points per decade,
and temperature $T$ from $10^{-3}$ to $10^5$\,eV with 10 points per decade.
The reference density $\rho_0$ was chosen to be that of the liquid cryogen;
it should be noted that the choice of this density is purely a convenience
in constructing a tabular EOS, where it is useful for the tabulation
to include the starting state in typical applications of the EOS model,
in order to reduce the sensitivity to interpolating functions.
The EOS were not adjusted to reproduce any empirical data.

As was found in the previous study \cite{Swift2018}, the 
electronic wavefunctions were computed reliably down to 10\,K or less
for densities corresponding to condensed matter, and to 100\,K or less for densities
down to 0.1\%\ of the ambient solid.
At lower densities, calculations completed successfully only for 
temperatures of several eV or more.
In contrast to the previous study,
calculations of the restoring force for infinitesimal displacements of the nucleus
gave an imaginary Einstein frequency, 
for densities slightly above that of the cryogenic liquid.
This behavior indicates the localization of electrons to an atom, 
and reflects the role of bonding orbitals and the van der Waals
interaction in stabilizing the liquid cryogen at low pressures.

The results of the atom-in-jellium calculations were, 
for each state of mass density $\rho$ and temperature $T$:
electronic contributions to the Helmholtz free energy $f$;
an ionic Einstein temperature $\theta_E$ and estimated Debye temperature 
$\theta_D$; 
the mean square displacement of the atom as a fraction of the Wigner-Seitz
radius $f_d$;
and the ionic contribution to $f$ using the generalized Debye model
with asymptotic ionic freedom \cite{Swift2018,Swift2019}.
The total electronic energy was used, including the ground state energy
at zero temperature:
this convention combines components sometimes presented as a 
cold compression curve and separate electron-thermal energy.
For isolated states where the atom-in-jellium calculation failed to converge,
polynomial interpolation from the surrounding states was used to complete
the EOS table.
For each state, the total Helmholtz free energy $f$ was calculated, and then 
differentiated using a quadratic fit to the three closest states in $\rho$ to
determine the pressure $p(\rho,T)$ in tabular form.
Similarly, quadratic fits in $T$ were differentiated to find the specific
entropy $s$, and hence the specific internal energy $e(\rho,T)$ in tabular form.
Together, these tables constitute an EOS
suitable for use in multi-physics hydrodynamic simulations.

\section{Principal Hugoniots for cryogenic liquids}
Although, without further processing, 
the atom-in-jellium EOS can be interrogated in compressed and heated states
where the calculations completed,
the ionic calculation was neither robust nor meaningful in cryogenic
liquid states.
This is a problem for their use in calculating the shock Hugoniot 
for an initially liquid sample, as a physical initial state is needed when
solving the Rankine-Hugoniot equations \cite{rh}.
An optimal solution to this problem would be to improve the EOS calculation to give a physical
ion-thermal energy, or to combine the calculation with a more reliable
technique for this region of state space.
However, this is a difficult problem in quantum chemistry,
outside the scope of the present work.
As long as our main interest is high-pressure states such as shocks into
the warm dense matter regime, a simpler approach is possible.

The Rankine-Hugoniot equations describe the conservation of mass,
momentum, and energy across a steady shock wave:
\begin{eqnarray}
\rho_0 u_s & = & \rho(u_s-u_p), \\
p_0+\rho_0 u_s^2 & = & p+\rho(u_s-u_p)^2, \\
\left[p_0+\rho_0e_0+\frac 12\rho_0 u_s^2\right]u_s & = & 
\left[p+\rho e+\frac 12\rho (u_s-u_p)^2\right](u_s-u_p)
\end{eqnarray}
where $u_s$ is the shock speed and $u_p$ the particle speed.
They are typically used to deduce the Hugoniot, or locus of states
accessible from a given initial state by the passage of a single, steady
shock wave, for matter described by an EOS of the form $p(\rho,e)$.
The initial state enters only through the quantities $\rho_0$, $e_0$, and $p_0$,
so the EOS need not be valid near the unshocked state if these quantities can
be obtained in a different way.
This aspect is used in calculations of detonating high explosives, where 
the EOS of the reaction products can be used without any physical representation
of the unshocked explosive \cite{he}.
For the present EOS, defined via the Helmholtz free energy,
the temperature $T$ was 
eliminated when solving the Rankine-Hugoniot equations \cite{rhnew}.

To deduce a Hugoniot from atom-in-jellium EOS with inaccurate or undefined
states around that of the initial cryogenic liquid, $(\rho_0,T_0)$, 
we first explored higher initial temperatures $T_0^\prime$
at the same mass density, until a usable state was found.
At this state $(\rho_0,T_0^\prime)$, the
specific internal energy from the atom-in-jellium calculation
took some value $e_0^\prime$.
A Hugoniot could be calculated from this initial state.
Next, we used a suitable previously-developed reference EOS 
to calculate a difference in
specific internal energy between the states:
\begin{equation}
\Delta e = e_{\mbox{ref}}(\rho_0,T_0^\prime)-e_{\mbox{ref}}(\rho_0,T_0).
\end{equation}
We then recalculated the atom-in-jellium Hugoniot, defining the initial
energy to be $e_0^\prime-\Delta e$.
This approach essentially uses the specific heat capacity from the reference EOS
to correct the atom-in-jellium state from a reliable value at $T_0^\prime$
to the desired $T_0$.

In the following comparisons for each element,
experimental shock data were taken from the Marsh and van Thiel compendia
\cite{Marsh1980,vanThiel1966}.
As well as the principal Hugoniot from the cryogenic liquid state,
we consider the cold compression curve, which explores higher densities,
and the isochore passing through the cryogenic liquid state,
which explores higher temperatures relevant to ablation
and expansion following a shock.

\subsection{Nitrogen}
\begin{figure}
\begin{center}\includegraphics[scale=0.72]{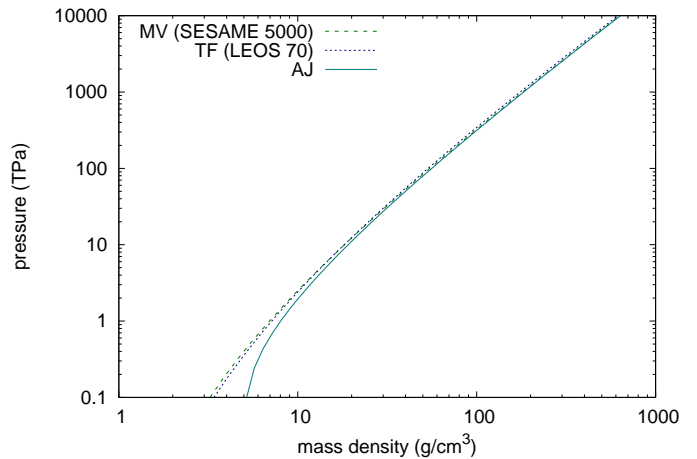}\end{center}
\caption{Cold compression curve for nitrogen,
   from molecular vibrations (MV) \cite{Kerley1980}, Thomas-Fermi (TF) \cite{leos},
   and the present atom-in-jellium results (AJ).}
\label{fig:nisothdp}
\end{figure}

\begin{figure}
\begin{center}\includegraphics[scale=0.72]{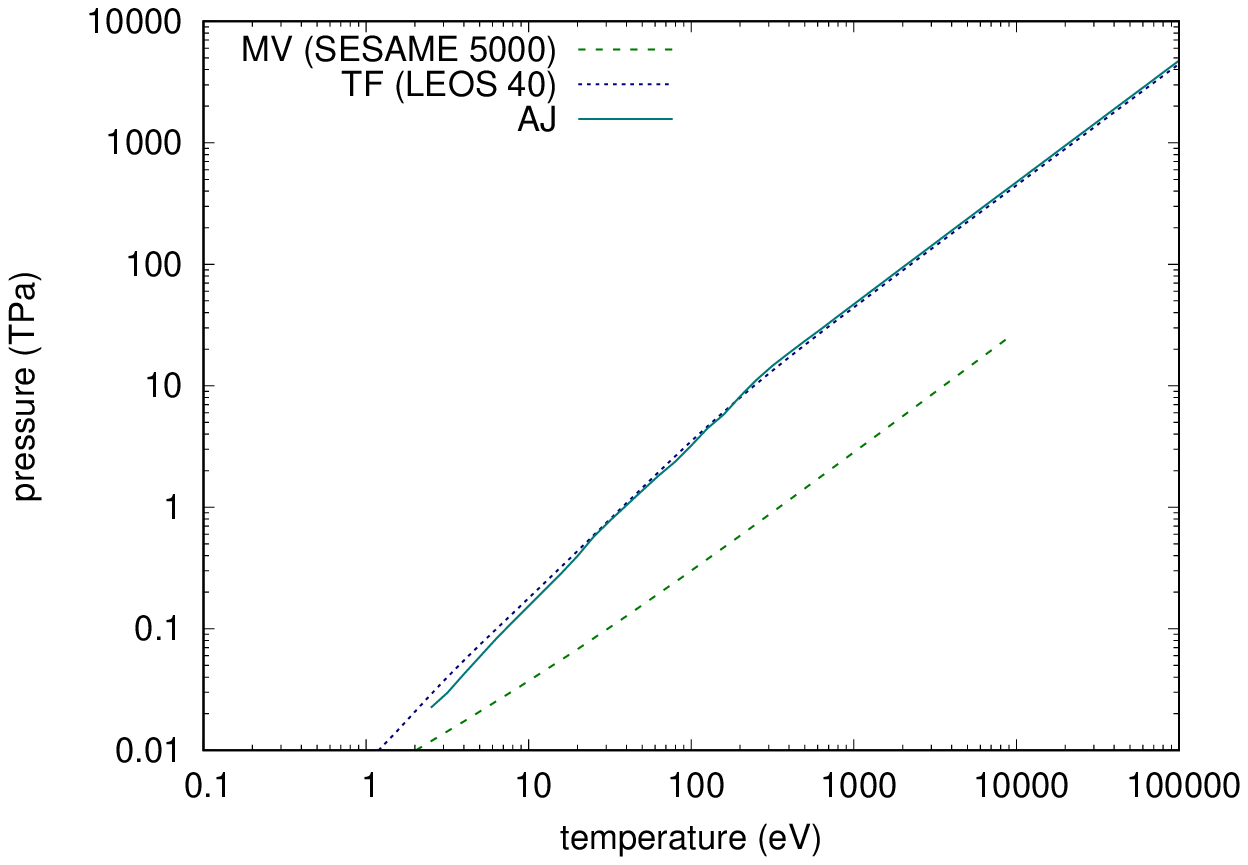}\end{center}
\caption{Principal isochore for nitrogen,
   from molecular vibrations (MV) \cite{Kerley1980}, Thomas-Fermi (TF) \cite{leos},
   and the present atom-in-jellium results (AJ).}
\label{fig:nisochtp}
\end{figure}

For nitrogen, apart from deviations around the initial state,
the atom-in-jellium calculations of the cold curve and principal isochore
lie close to those of the TF-based {\sc leos}~70 \cite{leos},
even down to the change in gradient around 300\,eV on the isochore.
Although the isochore from the MV-based {\sc sesame}~5000 \cite{Kerley1980}
falls well below that from the other EOS, its cold curve is remarkably consistent
with the TF-based model over the full range considered.
(Figs~\ref{fig:nisothdp} and \ref{fig:nisochtp}.)

\begin{figure}
\begin{center}\includegraphics[scale=0.72]{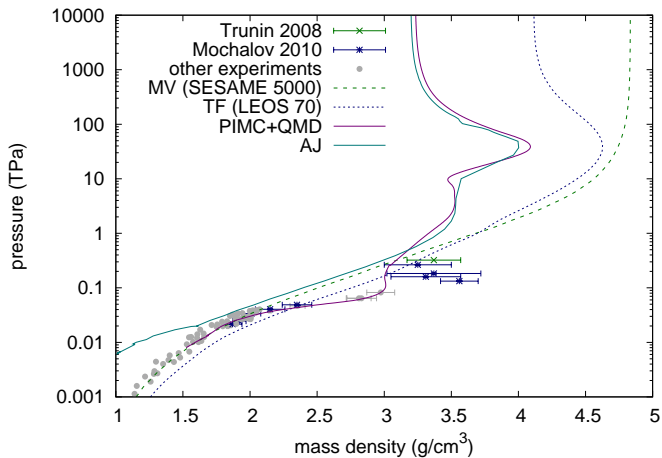}\end{center}
\caption{Shock Hugoniot for nitrogen,
   from molecular vibrations (MV) \cite{Kerley1980}, Thomas-Fermi (TF) \cite{leos},
   PIMC and QMD simulations \cite{Driver2016},
   shock measurements \cite{Zubarev1962,Dick1970,Nellis1980,Nellis1991,Trunin2008,Mochalov2010},
   and the present atom-in-jellium results (AJ).}
\label{fig:nhugdp}
\end{figure}

The atom-in-jellium Hugoniot passes through the experimental shock
measurements \cite{Zubarev1962,Dick1970,Nellis1980,Nellis1991,Trunin2008,Mochalov2010} 
around 50\,GPa.
At higher pressures, the Hugoniot is significantly stiffer than either 
the MV-based or the TF-based EOS.
The MV EOS omits dissociation and ionization, and shows a Gr\"uneisen-like
behavior where the shock density approaches a limit asymptotically.
The latter, TF-based, model
exhibits a peak compression at a similar pressure to the atom-in-jellium
EOS, except at a much higher density.
The atom-in-jellium Hugoniot also shows a clear feature corresponding to
ionization of the outer electrons around 3.5\,g/cm$^3$
The atom-in-jellium result follows QMD/PIMC predictions of the Hugoniot
\cite{Driver2016} much more closely, except that it fails to reproduce the
depression between 2 and 3\,g/cm$^3$ corresponding to dissociation
of the N$_2$ molecules.
The QMD Hugoniot reproduces this plateau well, though it predicts
the subsequent onset of stiffening at a lower density than
experiments using a spherically-converging shock wave
\cite{Trunin2008,Mochalov2010}.
(Fig.~\ref{fig:nhugdp}.)

\subsection{Oxygen}
\begin{figure}
\begin{center}\includegraphics[scale=0.72]{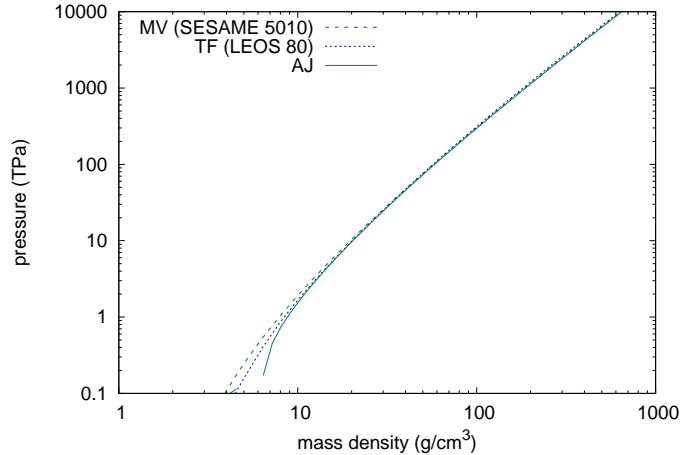}\end{center}
\caption{Cold compression curve for oxygen,
   from molecular vibrations (MV) \cite{Kerley1980}, Thomas-Fermi (TF) \cite{leos},
   and the present atom-in-jellium results (AJ).}
\label{fig:oisothdp}
\end{figure}

\begin{figure}
\begin{center}\includegraphics[scale=0.72]{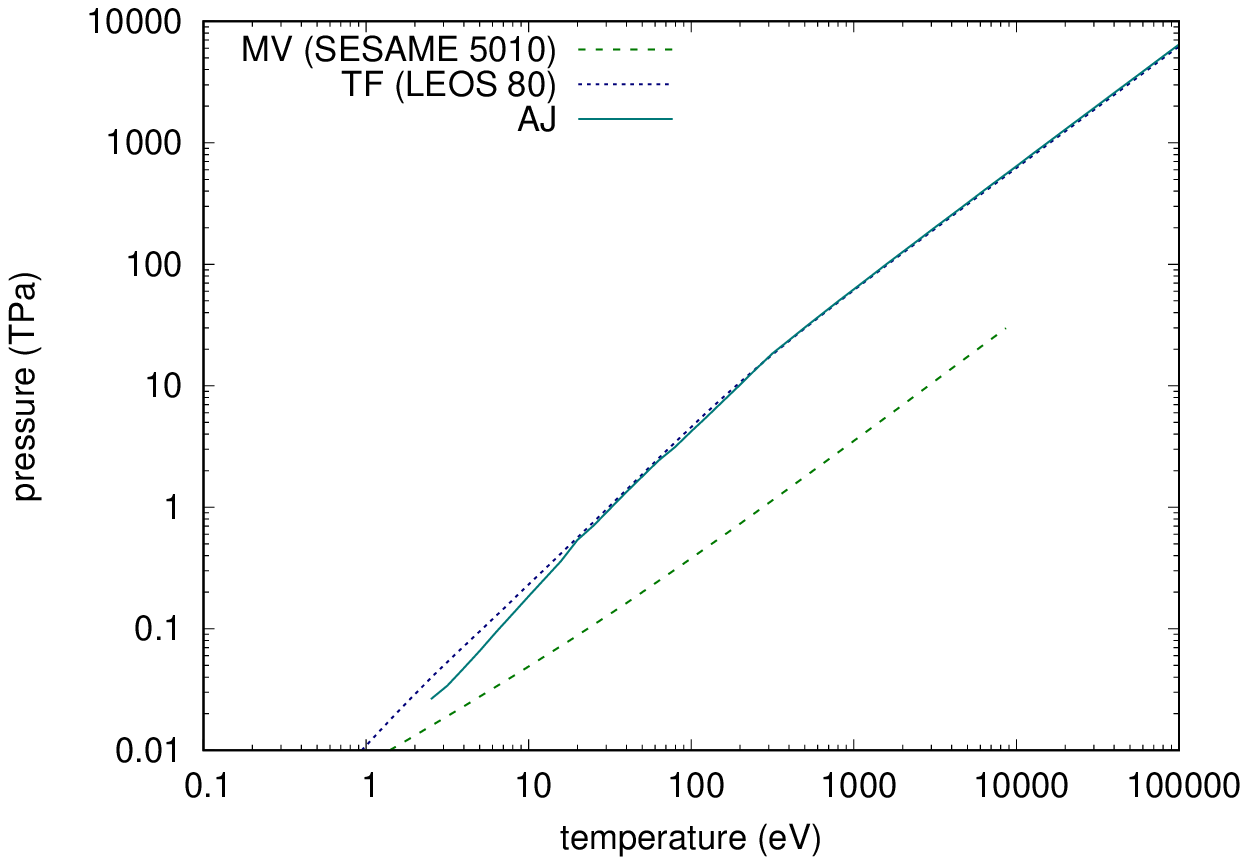}\end{center}
\caption{Principal isochore for oxygen,
   from molecular vibrations (MV) \cite{Kerley1980}, Thomas-Fermi (TF) \cite{leos},
   and the present atom-in-jellium results (AJ).}
\label{fig:oisochtp}
\end{figure}

For oxygen, comparisons of the cold curves and isochores are much the same
as for nitrogen.
Apart from deviations around the initial state,
the atom-in-jellium calculations 
lie close to those of the TF-based {\sc leos}~80 \cite{leos},
again replicating the change in gradient around 300\,eV on the isochore.
The isochore from the MV-based {\sc sesame}~5010 \cite{Kerley1980} again
falls well below that from the other EOS, but the cold curve is similar to
that from the TF-based EOS.
(Figs~\ref{fig:oisothdp} and \ref{fig:oisochtp}.)

\begin{figure}
\begin{center}\includegraphics[scale=0.72]{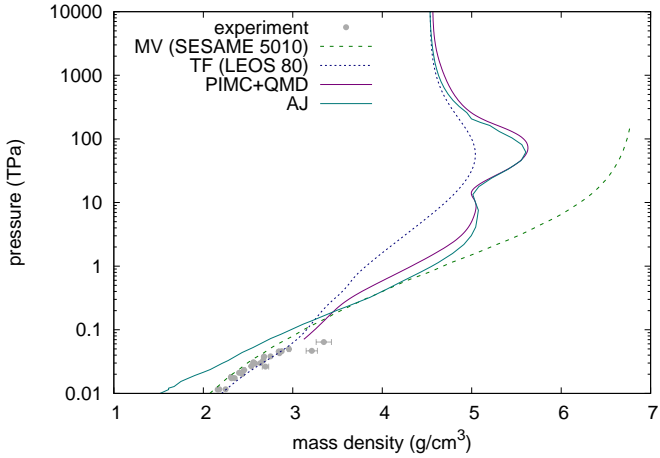}\end{center}
\caption{Shock Hugoniot for oxygen,
   from molecular vibrations (MV) \cite{Kerley1980}, Thomas-Fermi (TF) \cite{leos},
   PIMC and QMD simulations \cite{Driver2016},
   shock measurements \cite{Nellis1980,Hamilton1988},
   and the present atom-in-jellium results (AJ).}
\label{fig:ohugdp}
\end{figure}

For oxygen, the atom-in-jellium Hugoniot passes slightly above the 
experimental data, and then closely tracks the Hugoniot for the
MV-based EOS to 1.5\,TPa.
At higher pressures, it exhibits a peak compression at 50\,TPa, then
approaches the TF-based EOS for pressures above 1000\,TPa.
As with nitrogen, the atom-in-jellium calculations exhibit a distinct feature
around 5\,g/cm$^3$ from ionization of the outer electrons.
The maximum compression, although at a similar pressure to the TF-based EOS,
is at a significantly higher density.
As for nitrogen, the atom-in-jellium results
follow QMD/PIMC calculations of the Hugoniot \cite{Driver2016},
which in this case do not predict as strong of a dissociation feature.
In oxygen, the contribution of spin to the covalent bond is 
important \cite{Militzer2003}; the QMD calculations were spinless and thus
underpredict the dissociation feature.
Experimental data in the region of dissociation are relatively sparse,
but measurements \cite{Nellis1980,Hamilton1988} 
made consistently with those for nitrogen suggest a similar plateau.
(Fig.~\ref{fig:ohugdp}.)

\subsection{Fluorine}
\begin{figure}
\begin{center}\includegraphics[scale=0.72]{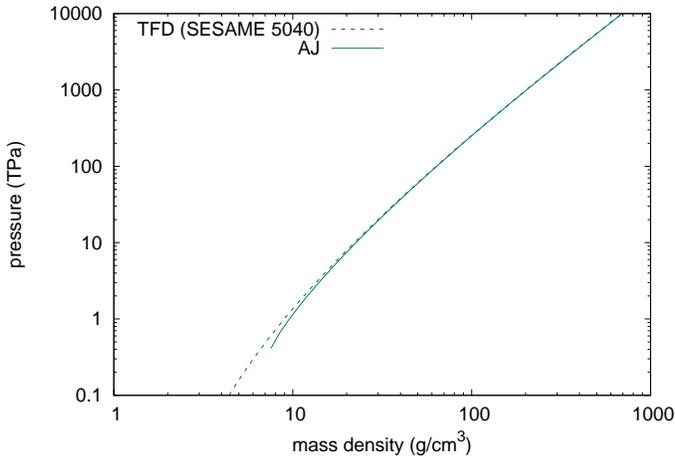}\end{center}
\caption{Cold compression curve for fluorine,
   from TFD \cite{Crockett2006} and the present atom-in-jellium (AJ) calculations.}
\label{fig:fisothdp}
\end{figure}

\begin{figure}
\begin{center}\includegraphics[scale=0.72]{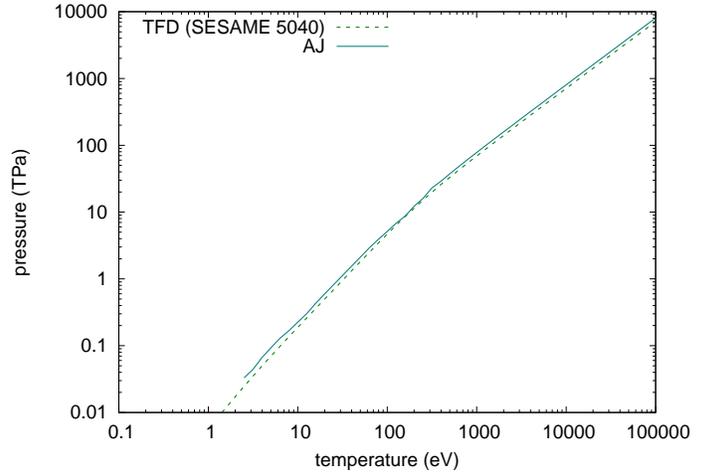}\end{center}
\caption{Principal isochore for fluorine,
   from TFD \cite{Crockett2006} and the present atom-in-jellium (AJ) calculations.}
\label{fig:fisochtp}
\end{figure}

For fluorine again, the atom-in-jellium cold curve and isochore
are similar to the corresponding curves from the TFD-based EOS
{\sc sesame} 5040 \cite{Crockett2006}, apart from deviations around
the initial state, and once more replicating the change in isochore
gradient around 300\,eV.
(Figs~\ref{fig:fisothdp} and \ref{fig:fisochtp}.)

\begin{figure}
\begin{center}\includegraphics[scale=0.72]{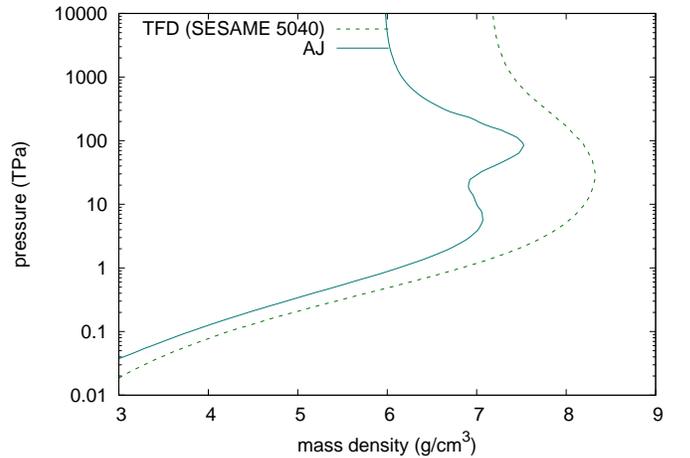}\end{center}
\caption{Shock Hugoniot for fluorine,
   from TFD \cite{Crockett2006} and the present atom-in-jellium (AJ) calculations.}
\label{fig:fhugdp}
\end{figure}

The atom-in-jellium Hugoniot again exhibits a distinct feature
around 7\,g/cm$^3$ corresponding to ionization of the outer electrons,
and a peak compression around 100\,TPa.
The peak compression is at a lower density than in the TFD-based EOS,
and more localized in pressure.
The atom-in-jellium EOS was constructed completely consistently with those
for nitrogen and oxygen, with no empirical parameters. 
We therefore deem it highly likely that
the Hugoniot from {\sc sesame} 5040 is up to $\sim$20\%\ too dense.
(Fig.~\ref{fig:fhugdp}.)

\section{Discussion}
It is instructive to compare the systematic behavior of the atom-in-jellium EOS
with previous, semi-empirical EOS for the Hugoniots of these cryogenic liquids.
Where experimental Hugoniot data are available, 
the atom-in-jellium EOS is inaccurate at low pressures,
as expected given its inability to capture the details of
interatomic forces, but becomes close to or coincident with shock data
at pressures of several tens to $O(100)$\,GPa.
As the shock pressure increases, the atom-in-jellium EOS was found to track
surprisingly close to the molecular EOS in the {\sc sesame} library,
deviating when effects from the excitation of electron shells became
significant and the molecular EOS would then be inaccurate.
As has been found for other elements, the atom-in-jellium Hugoniot exhibited
a relatively sharp peak in compression compared with the broader peak 
characteristic of a TF treatment of the electrons.
For the elements considered here, this feature originates from ionization
of the electrons in the $K$-shell.
Interestingly, the peak compression in the TF-based {\sc leos}
models varied non-systematically from the atom-in-jellium calculations:
higher for nitrogen but lower for oxygen, and with much greater difference
than was found previously for Be, Al, Si, Fe, and Mo
\cite{Swift2018}, which are solid at STP and more widely studied;
Al and Mo being regarded as standards for high pressure work.
Indeed, unlike these other elements, the TF calculation
of post-peak compression only matched the atom-in-jellium calculation for
oxygen.
Shock Hugoniots derived from different EOS models were found to differ
significantly, even when the cold curves and principal isochores were
much more similar:
the interplay between compression and heating in shock states can 
magnify the difference between EOS models.

In practice, because TF theory is inadequate around ambient
conditions, TF-based EOS are typically a combination of
an empirical Mie-Gr\"uneisen fit to properties at STP together with
shock and isothermal compression data, coupled to
TF theory at higher temperatures.
The markedly different behavior of shock Hugoniots derived from
TF-based EOS in comparison with our atom-in-jellium EOS
reflects the scarcity of shock data for these cryogenic liquids.
This observation highlights the value of the atom-in-jellium calculations:
although inadequate in the cryogenic liquid state,
their theoretical underpinnings are inherently more self-consistent,
so it is not necessary to make empirical adjustments to match shock data
above a few hundred gigapascals.
As such, they are likely to be more accurate in regimes where the EOS is not
constrained directly by experimental data,
particularly at elevated pressures and temperatures.
However, it is important to stress that atom-in-jellium calculations
do not describe molecular bonding, and thus the shock Hugoniots derived from
these EOS lack features associated with molecular dissociation.
The shock Hugoniot of fluorine should also exhibit this feature, 
though likely less pronounced than in oxygen,
as the trend should correlate with the energy of the covalent bond:
9.79, 5.15, and 1.63\,eV for N$_2$, O$_2$, and F$_2$ respectively, at STP.
Recent spinless QMD results reproduced the dissociation feature fairly well in
nitrogen but appeared to underpredict it in oxygen \cite{Driver2016}.

In accord with recent results from PIMC and QMD simulations where available,
our atom-in-jellium EOS predictions of the principal Hugoniots
of nitrogen, oxygen, and fluorine
indicate that EOS models for these elements currently in wide use 
have significant inaccuracies in the dense plasma regime,
densities of several grams per cubic centimeter and temperatures above a few 
electron-volts.

\section{Conclusions}
We have constructed wide-ranging EOS for nitrogen, oxygen, and fluorine
using the atom-in-jellium model, including ion-thermal and cold curve
energies as well as the electron-thermal energy for which atom-in-jellium
calculations have been used most widely.
Although the ion-thermal calculations are not valid around the
cryogenic liquid states for these elements, which are stabilized by
interatomic bonding and van der Waals forces not captured by the atom-in-jellium
model, it is possible to calculate shock Hugoniots from these initial states
by correcting preheated calculations to obtain a reasonable starting energy.

Apart from a regime affected by dissociation in N$_2$,
the atom-in-jellium Hugoniots match predictions based on QMD and PIMC 
where they are available.

As was found with elements which are solid at STP, 
despite disagreements at low pressures, 
shock Hugoniots derived from the atom-in-jellium EOS are close to or match
shock data at pressures approaching around 100\,GPa.
At higher pressures, where no shock data currently exist, there are substantial
differences between the Hugoniots from atom-in-jellium EOS and those from
other EOS using a Thomas-Fermi treatment of the electrons or neglecting
dissociation and ionization.
As the atom-in-jellium EOS were calculated self-consistently, without any
empirical adjustment to match data for any of these elements, this deviation
is likely to reflect inaccuracies in the previous EOS models.
Such inaccuracies imply that EOS for planetary materials such
as oxides and silicates, constructed in the warm dense matter regime by
mixing previous elemental EOS such as these, 
may well be inaccurate for applications such as
the internal structure of giant planets and the dynamics of planetary collisions.

\section*{Acknowledgments}
Kevin Driver kindly provided results from his
PIMC and QMD simulations.

The work at LLNL was performed under the auspices of
the U.S. Department of Energy under contract DE-AC52-07NA27344.
B.M. was supported by the U.S. Department of Energy 
(Grant DE-SC0016248) and by the University of California through the
multi-campus research Award 0013725.


\begin{thebibliography}{10}
\bibitem{sesame}{S.P.~Lyon and J.D.~Johnson, Los Alamos National Laboratory
   report LA-UR-92-3407 (1992).}
\bibitem{leos}{R.M.~More, K.H.~Warren, D.A.~Young and G.B.~Zimmerman,
   Phys. Fluids {\bf 31}, 3059 (1988);
   D.A.~Young and E.M.~Corey, J.~Appl. Phys. {\bf 78}, 3748 (1995).}
\bibitem{Kerley1980}{G.~Kerley and J.~Abdallah,
   J.~Chem. Phys. {\bf 73}, 5337 (1980).}
\bibitem{Crockett2006}{S.D.~Crockett, Los Alamos National Laboratory report
   LA-UR-06-8404 (2006).}
\bibitem{pimc}{E.L.~Pollock and D.M.~Ceperley, Phys. Rev.~B {\bf 30}, 2555 (1984).}
\bibitem{qmd}{
   L.~Collins, I.~Kwon, J.~Kress, N.~Troullier, and D.~Lynch,
   Phys. Rev.~E {\bf 52}, 6202 (1995).
   }
\bibitem{Liberman1979}{D.A.~Liberman, Phys. Rev.~B {\bf 20}, 12, 4981 (1979).}
\bibitem{Benedict2014}{L.X.~Benedict, K.P.~Driver, S.~Hamel, B.~Militzer, T.~Qi, A.A.~Correa, A.~Saul, and E.~Schwegler, Phys. Rev. B {\bf 89}, 224109 (2014).}
\bibitem{Driver2017}{K.P.~Driver and B.~Militzer, Phys. Rev. E {\bf 95}, 043205 (2017).}
\bibitem{Liberman1990}{D.A.~Liberman and B.I.~Bennett, Phys. Rev.~B {\bf 42}, 2475 (1990).}
\bibitem{Swift2018}{D.C.~Swift, T.~Lockard, R.G.~Kraus, L.X.~Benedict,
   P.A.~Sterne,M.~Bethkenhagen,  S.~Hamel, and B.I.~Bennett,
   Phys. Rev. E {\bf 99}, 063210 (2019).} 
\bibitem{Swift2019}{D.C.~Swift, T.~Lockard, M.~Bethkenhagen, S.~Hamel,
   A.~Correa, L.X.~Benedict, P.A.~Sterne, and B.I.~Bennett,
   submitted and {\tt arXiv:1905.08911} (2019).}
\bibitem{Driver2016}{K.P.~Driver and B.~Militzer,
   Phys. Rev. B {\bf 93}, 064101 (2016).}
\bibitem{Zubarev1962}{V.N.~Zubarev and G.S.~Telegin, 
   Sov. Phys. Dokl. {\bf 7}, 34 (1962).}
\bibitem{Dick1970}{R.D.~Dick, J.~Chem. Phys. {\bf 52}, 6021 (1970).}
\bibitem{Nellis1980}{W.J.~Nellis and A.C.~Mitchell,
   J. Chem. Phys. {\bf 73}, 6137 (1980).}
\bibitem{Nellis1991}{W.J.~Nellis, H.B.~Radousky, D.C.~Hamilton, A.C.~Mitchell, N.C.~Holmes, K.B.~Christianson, and M.~van~Thiel,
   J. Chem. Phys. {\bf 93}, 2244 (1991).}
\bibitem{Trunin2008}{R.F.~Trunin, G.V.~Boriskov, A.I.~Bykov, A.B.~Medvedev, G.V.~Simakov, and A.N.~Shuikin,
   J.~Exp. Theor. Phys. Lett., {\bf 88}, 3, pp~189–191 (2008).}
\bibitem{Mochalov2010}{M.A.~Mochalov, M.V.~Zhernokletov, R.I.~Il’kaev, A.L.~Mikhailov, V.E.~Fortov, V.K.~Gryaznov, I.L.~Iosilevskiy, A.B.~Mezhevov, A.E.~Kovalev, S.I.~Kirshanov, Yu.A.~Grigor’eva, M.G.~Novikov, and A.N.~Shuikin,
   J.~Exp. Theor. Phys. {\bf 110}, 1, pp~67–80 (2010).}
\bibitem{Hamilton1988}{D.C.~Hamilton, W.J.~Nellis, A.C.~Mitchell, F.H.~Ree, and M.~van Thiel,
   J.~Chem. Phys. {\bf 88}, 5042 (1988).}
\bibitem{tf}{L.H.~Thomas,
   Proc. Cambridge Phil. Soc. {\bf 23}, 5, 542–548 (1927);
   E.~Fermi, 
   Rend. Accad. Naz. Lincei. {\bf 6}, 602–607 (1927).}
\bibitem{rh}{M.R.~Boslough and J.R.~Asay, in
   J.R.~Asay, M.~Shahinpoor (Eds),
   ``High-Pressure Shock Compression of Solids''
   (Springer-Verlag, New York, 1992).}
\bibitem{he}{W.~Fickett and W.C.~Davis,
   ``Detonation'' (University of California, Berkeley, 1979).}
\bibitem{rhnew}{D.C.~Swift and M.~Millot, in preparation.}
\bibitem{Marsh1980}{S.P.~Marsh (Ed), {\it LASL Shock Hugoniot Data}
   (University of California, Berkeley, 1980).}
\bibitem{vanThiel1966}{M.~van Thiel,
   {\it Compendium of Shock Wave Data},
   Lawrence Livermore National Laboratory report UCRL-50108 (1966).}
\bibitem{Militzer2003}{B.~Militzer, F.~Gygi, and G.~Galli,
   Phys. Rev. Lett. {\bf 91}, 265503 (2003).}
\end{thebibliography}
\end{document}